\documentclass[10pt,twocolumn,letterpaper]{article}

\usepackage{setspace}
\usepackage{cvpr}
\usepackage{times}
\usepackage{epsfig}
\usepackage{graphicx}
\usepackage{amsmath}
\usepackage{amssymb}
\usepackage{bm}
\usepackage{booktabs}
\usepackage{array}
\usepackage{xspace}
\usepackage{multirow}
\usepackage{color}
\usepackage{placeins}
\usepackage{enumitem}
\usepackage{xcolor,colortbl}
\usepackage{xspace}

\usepackage[font={small}]{caption}
\usepackage[pagebackref=true,breaklinks=true,letterpaper=true,colorlinks,bookmarks=false]{hyperref}

\cvprfinalcopy

\newcommand\blfootnote[1]{%
  \begingroup
  \renewcommand\thefootnote{}\footnote{#1}%
  \addtocounter{footnote}{-1}%
  \endgroup
}

\definecolor{tablegreen}{rgb}{0.09, 0.45, 0.27}
\definecolor{tablered}{rgb}{0.55, 0.0, 0.0}
\definecolor{lightgray}{rgb}{0.33, 0.41, 0.47}
\newcommand{\name}{L3C\xspace}

\newcommand{\lossidx}{i}
\newcommand{\lossparts}[1]{F^{(#1)}_\lossidx}

\newcommand{\expectE}{\mathbb{E}}

\newcommand{\Preal}{\tilde{p}}
\newcommand{\zunquantized}{z'}
\newcommand{\R}{\mathbb{R}}
\newcommand{\levels}{\mathbb{L}}
\newcommand{\level}{\ell}
\newcommand{\sigmaq}{\sigma_q}
\newcommand{\X}{\mathcal{X}}

\newcommand{\Bicubic}{\mathcal{B}}

\newcommand{\sigmoid}{\text{sigmoid}}

\newcommand{\fs}[1]{f^{(#1)}}

\newcommand{\pmix}{p_m}
\newcommand{\plogistic}{p_l}

\newcommand{\tablenote}[1]{$^{\textcolor{lightgray}{#1}}$}
\newcommand{\deftablenote}[1]{$\textcolor{lightgray}{\small [#1]}$}

\newcommand{\raisek}{RAISE-1k\xspace}
\newcommand{\jpegk}{JPEG\kern0.2ex2000\xspace}

\definecolor{Gray}{gray}{0.9}

\makeatletter
\renewcommand{\paragraph}{%
  \@startsection{paragraph}{4}%
  {\z@}{2ex \@plus 1.25ex \@minus .5ex}{-1em}%
  {\normalfont\normalsize\bfseries}%
}
\makeatother

\newcolumntype{P}[1]{>{\centering\arraybackslash}p{#1}}

\begin{document}

\title{Practical Full Resolution Learned Lossless Image Compression}

\author{
    \begin{tabular}{ccccc}
        Fabian Mentzer &
        Eirikur Agustsson &
        Michael Tschannen &
        Radu Timofte &
        Luc Van Gool\\
        {\scriptsize mentzerf@vision.ee.ethz.ch} &
        {\scriptsize aeirikur@vision.ee.ethz.ch} &
        {\scriptsize michaelt@nari.ee.ethz.ch} &
        {\scriptsize timofter@vision.ee.ethz.ch} &
        {\scriptsize vangool@vision.ee.ethz.ch} 
    \end{tabular}\\[3ex]
    ETH Z\"urich, Switzerland\vspace{-1ex}
}

\maketitle

\begin{abstract}
We propose the first practical learned lossless image compression system, L3C, and show that it outperforms the popular engineered codecs, PNG, WebP and \jpegk.
At the core of our method is a fully parallelizable hierarchical probabilistic model for adaptive entropy coding which is optimized end-to-end for the compression task.
In contrast to recent autoregressive discrete probabilistic models such as PixelCNN, our method i) models the image distribution jointly with learned auxiliary representations instead of exclusively modeling the image distribution in RGB space, and ii) only requires three forward-passes to predict all pixel probabilities instead of one for each pixel.
As a result, L3C obtains over two orders of magnitude speedups when sampling compared to the fastest PixelCNN variant (Multiscale-PixelCNN). 
Furthermore, we find that learning the auxiliary representation is crucial and outperforms predefined auxiliary representations such as an RGB pyramid significantly.
\end{abstract}

\vspace{-3ex}
\section{Introduction}
\blfootnote{\textbf{Version 3}: This is an updated version of the paper. See suppl.~\ref{sec:version3}.}
Since likelihood-based discrete generative models learn a probability distribution over pixels, they can in theory be used for lossless image compression \cite{theis2016note}. However, recent work on learned compression using deep neural networks has solely focused on lossy compression \cite{balle2016end, theis2017lossy, toderici2016full, rippel17a, agustsson2017soft,agustsson2018generative,tschannen2018lossy}. Indeed, the literature on discrete generative models~\cite{van2016pixel, van2016conditional, Salimans2017pcnnpp, reed2017parallel, kolesnikov2017pixelcnn} has largely ignored the application as a lossless compression system, with neither bitrates nor runtimes being compared with classical codecs such as PNG~\cite{pngurl}, WebP~\cite{webpurl}, \jpegk~\cite{skodras2001jpeg2000}, and FLIF~\cite{flif2016}.
This is not surprising as (lossless) entropy coding using likelihood-based discrete generative models amounts to a decoding complexity essentially identical to the sampling complexity of the model, which renders many of the recent state-of-the-art autoregressive models such as PixelCNN~\cite{van2016pixel}, PixelCNN++~\cite{Salimans2017pcnnpp}, and Multiscale-PixelCNN~\cite{reed2017parallel} impractical, requiring minutes or hours on a GPU to generate moderately large images, typically ${<}256 \times 256$px (see Table~\ref{table:times}). 
The computational complexity of these models is mainly caused by the sequential nature of the sampling (and thereby decoding) operation, where a forward pass needs to be computed for every single (sub) pixel of the image in a raster scan order.

\begin{figure}[t!]
\centering
\includegraphics[width=\linewidth]{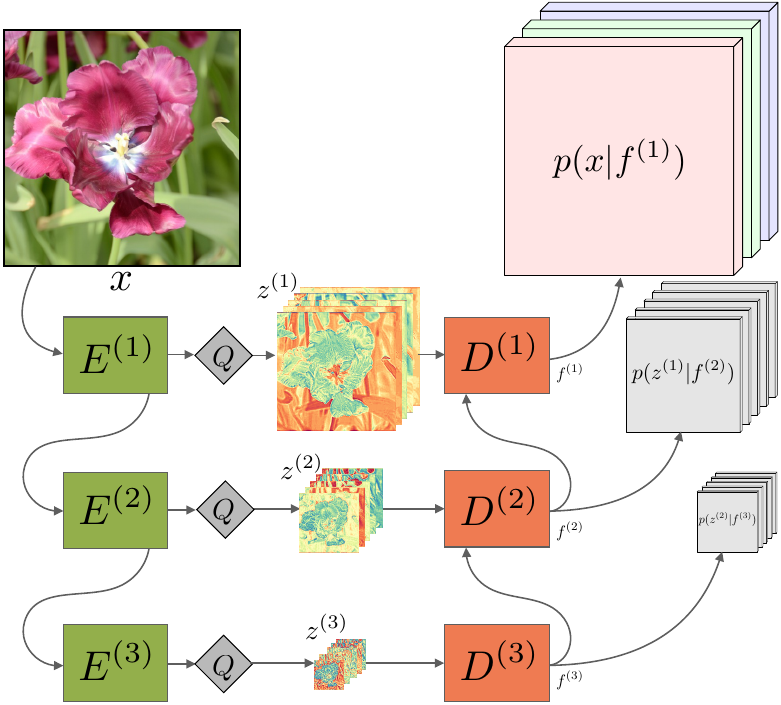}
    \caption{\label{fig:arch}Overview of the architecture of \name. The feature extractors $E^{(s)}$ compute quantized (by $Q$) auxiliary hierarchical feature representation $z^{(1)}, \ldots, z^{(S)}$ whose joint distribution with the image $x$, $p(x,z^{(1)}, \ldots, z^{(S)})$, is modeled using the non-autoregressive predictors $D^{(s)}$. The features $f^{(s)}$ summarize the information up to scale $s$ and are used to predict $p$ for the next scale.}
\end{figure}

In this paper, we address these challenges and develop a fully parallelizeable learned lossless compression system, outperforming the popular classical systems PNG, WebP and \jpegk.

Our system (see Fig.~\ref{fig:arch} for an overview) is based on a hierarchy of fully parallel learned feature extractors and predictors which are trained jointly for the compression task. Our code is available online\footnote{\label{fn:github}\url{github.com/fab-jul/L3C-PyTorch}}. The role of the feature extractors is to build an auxiliary hierarchical feature representation which helps the predictors to model both the image and the auxiliary features themselves. Our experiments show that learning the feature representations is crucial, and heuristic (predefined) choices such as a multiscale RGB pyramid lead to suboptimal performance.

In more detail, to encode an image $x$, we feed it through the $S$ feature extractors $E^{(s)}$ and predictors $D^{(s)}$. Then, we obtain the predictions of the probability distributions $p$, for both $x$ and the auxiliary features $z^{(s)}$, in parallel in a single forward pass. These predictions are then used with an adaptive arithmetic encoder to obtain a compressed bitstream of both $x$ and the auxiliary features (Sec.~\ref{sec:losslesscomp} provides an introduction to arithmetic coding). However, the arithmetic decoder now needs $p$ to be able to decode the bitstream. Starting from the lowest scale of auxiliary features $z^{(S)}$, for which we assume a uniform prior, $D^{(S)}$ obtains a prediction of the distribution of the auxiliary features of the next scale, $z^{(S-1)}$, and can thus decode them from the bitstream. Prediction and decoding is alternated until the arithmetic decoder obtains the image $x$. 
The steps are visualized in Fig.~\ref{fig:encdecdetails}.

In practice, we only need to use $S=3$ feature extractors and predictors for our model, so when decoding we only need to perform three parallel (over pixels) forward passes in combination with the adaptive arithmetic coding.

The parallel nature of our model enables it to be orders of magnitude faster for decoding than autoregressive models, while learning enables us to obtain compression rates competitive with state-of-the-art engineered lossless codecs.

In summary, our contributions are the following:
\begin{itemize}[leftmargin=*]
    \item We propose a fully parallel hierarchical probabilistic model, learning both the feature extractors that produce an auxiliary feature representation to help the prediction task, as well as the predictors which model the joint distribution of all variables (Sec.~\ref{sec:method}).
    \item We show that entropy coding based on our \emph{non-autoregressive} probabilistic model optimized for discrete log-likelihood can obtain compression rates outperforming WebP, \jpegk and PNG, the latter by a large margin. We are only marginally outperformed by the state-of-the-art, FLIF, while being conceptually much simpler (Sec.~\ref{sec:results_compression}).
    \item At the same time, our model is practical in terms of runtime complexity and orders of magnitude faster than PixelCNN-based approaches. In particular, our model is \emph{$5.31\cdot10^4\times$ faster than PixelCNN++}\cite{Salimans2017pcnnpp} and \emph{$5.06\cdot10^2\times$ faster than the highly speed-optimized MS-PixelCNN}~\cite{reed2017parallel} 
    (Sec.~\ref{sec:results_pixelcnn}).
\end{itemize}

\section{Related Work} \label{sec:related_work}

\paragraph{Likelihood-Based Generative Models} As previously mentioned, essentially all likelihood-based discrete generative models can be used with an arithmetic coder for lossless compression. A prominent group of models that obtain state-of-the-art performance are variants of the auto-regressive PixelRNN/PixelCNN \cite{van2016pixel, van2016conditional}. PixelRNN and PixelCNN organize the pixels of the image distribution as a sequence and predict the distribution of each pixel conditionally on (all) previous pixels using an RNN and a CNN with masked convolutions, respectively. These models hence require a number of network evaluations equal to the number of predicted sub-pixels\footnote{\label{fn:subpixel}A RGB ``pixel'' has 3 ``sub-pixels'', one in each channel.} ($3\cdot\! W\!\cdot\! H$).
PixelCNN++~\cite{Salimans2017pcnnpp} improves on this in various ways, including modeling the joint distribution of each pixel, thereby eliminating conditioning on previous channels and reducing to $W\!\cdot\! H$ forward passes. 
MS-PixelCNN \cite{reed2017parallel} parallelizes PixelCNN by reducing dependencies between blocks of pixels and processing them in parallel with shallow PixelCNNs, requiring~$O(\log WH)$ forward passes. \cite{kolesnikov2017pixelcnn} equips PixelCNN with auxiliary variables (grayscale version of the image or RGB pyramid) to encourage modeling of high-level features, thereby improving the overall modeling performance. \cite{chen2017pixelsnail, parmar2018image} propose autoregressive models similar to PixelCNN/PixelRNN, but they additionally rely on attention mechanisms to increase the receptive field.

\paragraph{Engineered Codecs} The well-known \emph{PNG}~\cite{pngurl} operates in two stages: first the image is reversibly transformed to a more compressible representation with a simple autoregressive filter that updates pixels based on surrounding pixels, then it is compressed with the deflate algorithm~\cite{deutsch1996deflate}.
\emph{WebP}~\cite{webpurl} uses more involved transformations, including the use of entire image fragments to encode new pixels and a custom entropy coding scheme.
\emph{\jpegk}~\cite{skodras2001jpeg2000} includes a lossless mode where tiles are reversibly transformed before the coding step, instead of irreversibly removing frequencies. 
The current state-of-the-art (non-learned) algorithm is \emph{FLIF}~\cite{flif2016}. It relies on powerful preprocessing and a sophisticated entropy coding method based on CABAC~\cite{richardson2004h} called MANIAC, which grows a dynamic decision tree per channel as an adaptive context model during encoding.

\paragraph{Context Models in Lossy Compression} In lossy compression, context models have been studied as a way to efficiently losslessly encode the obtained image representations. Classical approaches are discussed in~\cite{context5,context2,context4,context3,context1}. Recent learned approaches include~\cite{li2017learning, mentzer2018cvpr, minnen2018joint}, where shallow autoregressive models over latents are learned.
\cite{balle2018variational}~presents a model somewhat similar to \name: Their autoencoder is similar to our fist scale, and the hyper encoder/decoder is similar to our second scale. However, since they train for lossy image compression, their autoencoder predicts RGB pixels directly. Also, they predict uncertainties $\sigma$ for $z^{(1)}$ instead of a mixture of logistics. Finally, instead of learning a probability distribution for $z^{(2)}$, they assume the entries to be i.i.d.\ and fit a univariate non-parametric density model, whereas in our model, many more stages can be trained and applied recursively.

\paragraph{Continuous Likelihood Models for Compression}
The objective of continuous likelihood models, such as VAEs \cite{kingma2013auto} and RealNVP \cite{dinh2016density}, where $p(x')$ is a continuous distribution, is closely related to its discrete counterpart. In particular, by setting $x'=x+u$ where $x$ is the discrete image and $u$ is uniform quantization noise, the continuous likelihood of $p(x')$  is a lower bound on the likelihood of the discrete $q(x)= \mathbb E_u[p(x')]$ \cite{theis2016note}. However, there are two challenges for deploying such models for compression. First, the discrete likelihood $q(x)$ needs to be available (which involves a non-trivial integration step). Additionally, the memory complexity of (adaptive) arithmetic coding depends on the size of the domain of the variables of the factorization of $q$ (see Sec.~\ref{sec:losslesscomp} on (adaptive) arithmetic coding). Since the domain grows exponentially in the number of pixels in $x$, unless $q$ is factorizable, it is not feasible to use it with adaptive arithmetic coding.
\section{Method} \label{sec:method}

\subsection{Lossless Compression} \label{sec:losslesscomp}

In general, in lossless compression, some stream of symbols $x$ is given, which are drawn independently and identically distributed (i.i.d.)\ from a set $\X = \{1, \dots, |\X|\}$ according to the probability mass function $\Preal$. The goal is to encode this stream into a bitstream of minimal length using a ``code'', s.t.\ a receiver can decode the symbols from the bitstream. Ideally, an encoder minimizes the expected bits per symbol $\tilde L = \sum_{j \in \X} \Preal(j) \ell(j)$, where $\ell(j)$ is the length of encoding symbol $j$ (i.e., more probable symbols should obtain shorter codes). Information theory provides (e.g.,~\cite{cover2012elements}) the bound $\tilde L \ge H(\Preal)$ for \emph{any} possible code, where $H(\Preal) = \expectE_{j \sim \Preal}[-\log \Preal(j)]$ is the Shannon entropy~\cite{shannon1948it}.
    
\paragraph{Arithmetic Coding} A strategy that almost achieves the lower bound $H(\Preal)$ (for long enough symbol streams) is arithmetic coding~\cite{witten1987arithmetic}.\footnote{We use (adaptive) arithmetic coding for simplicity of exposition, but any adaptive entropy-achieving coder can be used with our method.} It encodes the entire stream into a single number $a' \in [0, 1)$, by subdividing $[0, 1)$ in each step (encoding one symbol) as follows: Let $a, b$ be the bounds of the current step (initialized to $a=0$ and $b=1$ for the initial interval $[0,1)$). We divide the interval $[a, b)$ into $|\X|$ sections where the length of the $j$-th section is $\Preal(j) / (b-a)$. Then we pick the interval corresponding to the current symbol, i.e., we update $a, b$ to be the boundaries of this interval. We proceed recursively until no symbols are left. Finally, we transmit $a'$, which is $a$ rounded to the smallest number of bits s.t.\ $a' \ge a$. Receiving $a'$ together with the knowledge of the number of encoded symbols and $\Preal$ uniquely specifies the stream and allows the receiver to decode.

\paragraph{Adaptive Arithmetic Coding} In contrast to the i.i.d.\ setting we just described, in this paper we are interested in losslessly encoding the pixels of a natural image, which are known to be heavily correlated and hence not i.i.d.\ at all. Let $x_t$ be the sub-pixels\textsuperscript{\ref{fn:subpixel}} of an image $x$, and $\Preal_\text{img}(x)$ the joint distribution of all sub-pixels. We can then consider the factorization $\Preal_\text{img}(x) = \prod_t \Preal(x_t | x_{t-1}, \dots, x_1)$. 
Now, to encode $x$, we can consider the sub-pixels $x_t$ as our symbol stream and encode the $t$-th symbol/sub-pixel using $\Preal(x_t | x_{t-1}, \dots, x_1)$. Note that this corresponds to varying the $\Preal(j)$ of the previous paragraph during encoding, and is in general referred to as \emph{adaptive} arithmetic coding (AAC)~\cite{witten1987arithmetic}. For AAC the receiver also needs to know the varying $\Preal$ at every step, i.e., they must either be known a priori or the factorization must be causal (as above) so that the receiver can calculate them from already decoded symbols. 

\paragraph{Cross-Entropy} In practice, the exact $\Preal$ is usually unknown, and instead is estimated by a model $p$. Thus, instead of using length $\log 1/\Preal(x)$ to encode a symbol $x$, we use the sub-optimal length $\log 1/p(x)$. 
Then
\begin{align}
H(\Preal, p) 
&= \expectE_{j \sim \Preal}\left[-\log p(j)\right] \nonumber \\
    &= -\sum_{j \in \X} \Preal(j) \log p(j) \label{eq:crossent}
\end{align}
is the resulting expected (sub-optimal) bits per symbol, and is called \textbf{cross-entropy}~\cite{cover2012elements}.

Thus, given some $p$, we can minimize the bitcost needed to encode a symbol stream with symbols distributed according to $\Preal$ by minimizing Eq.~\eqref{eq:crossent}. This naturally generalizes to the non-i.i.d.\ case described in the previous paragraph by using different $\Preal(x_t)$ and $p(x_t)$ for each symbol $x_t$ and minimizing $\sum_t H(\Preal(x_t), p(x_t))$.

The following sections describe how a hierarchical causal factorization of $p_\text{img}$ for natural images can be used to efficiently do learned lossless image compression (\name).

\subsection{Architecture} \label{sec:arch}

\begin{figure*}[ht!]
\centering
\includegraphics[width=\textwidth]{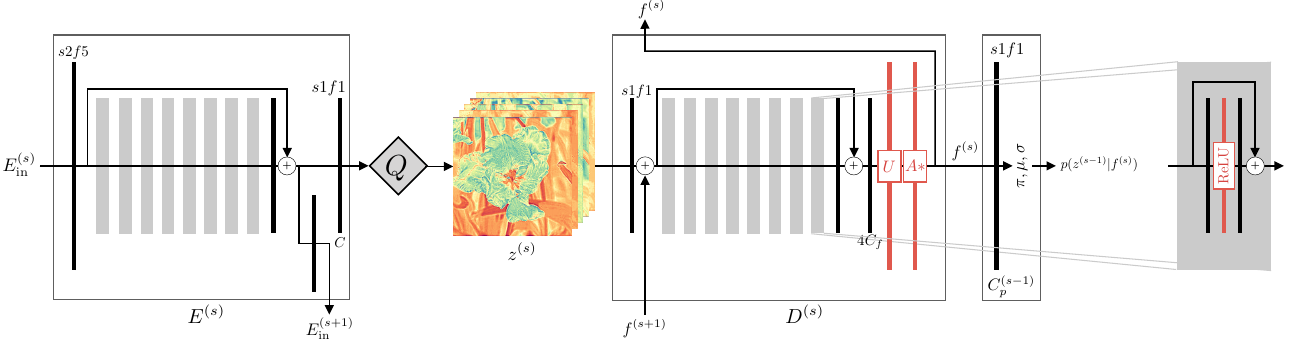}
    \caption{Architecture details for a single scale $s$. For $s = 1$, $E_\text{in}^{(1)}$ is the RGB image $x$ normalized to $[-1, 1]$. All vertical \textbf{black} lines are convolutions, which have $C_f=64$ filters, except when denoted otherwise beneath. The convolutions are stride 1 with $3{\times}3$ filters, except when denoted otherwise above (using $sSf\kern-1.5ptF$ = stride $s$, filter $f$). We add the features $f^{(s+1)}$ from the predictor $D^{(s+1)}$ to those of the first layer of $D^{(s)}$ (a skip connection between scales). The gray blocks are residual blocks, shown once on the right side.
    $C$ is the number of channels of $z^{(s)}$, $C_p^{(s-1)}$ is the final number of channels, see Sec.~\ref{sec:mixture_model}.
    Special blocks are denoted in red: $U$ is pixelshuffling upsampling~\cite{Shi2016RealTimeSI}. $A*$ is the ``atrous convolution'' layer described in Sec.~\ref{sec:arch}.
    We use a heatmap to visualize $z^{(s)}$, see Sec.~\ref{sec:visualize_repr}.
    }
\label{fig:arch_detail}
\end{figure*}

A high-level overview of the architecture is given in Fig.~\ref{fig:arch}, while Fig.~\ref{fig:arch_detail} shows a detailed description for one scale $s$.
Unlike autoregressive models such as PixelCNN and PixelRNN, which factorize the image distribution autoregressively over sub-pixels $x_t$ as $p(x) = \prod_{t=1}^T p(x_t|x_{t-1}, \ldots, x_1)$, we jointy model all the sub-pixels and introduce a learned hierarchy of auxiliary feature representations $z^{(1)}, \ldots, z^{(S)}$ to simplify the modeling task.We fix the dimensions of $z^{(s)}$ to be $C{\times}H'{\times}W'$, where the number of channels $C$ is a hyperparameter ($C=5$ in our reported models), and $H'=H/2^s, W'=W/2^s$ given a $H{\times}W$-dimensional image.\footnote{Considering that $z^{(s)}$ is quantized, this conveniently upper bounds the information that can be contained within each $z^{(s)}$, however, other dimensions could be explored.}
Specifically, we model the joint distribution of the image $x$ and the feature representations $z^{(s)}$ as
\begin{align*}
    p(x, z^{(1)}, &\ldots, z^{(S)}) = \\[-2ex]
    &p(x|z^{(1)}, \ldots, z^{(S)}) \prod_{s=1}^{S} p(z^{(s)}|z^{(s+1)}, \ldots, z^{(S)})
\end{align*}
where $p(z^{(S)})$ is a uniform distribution. The feature representations can be hand designed or learned. Specifically, on one side, we consider an RGB pyramid with $z^{(s)}=\Bicubic_{2^s}(x)$, where $\Bicubic_{2^s}$ is the bicubic (spatial) subsampling operator with subsampling factor $2^s$. On the other side, we consider a learned representation $z^{(s)}=F^{(s)}(x)$ using a feature extractor $F^{(s)}$. We use the hierarchical model shown in Fig.~\ref{fig:arch} using the composition 
$F^{(s)} = Q \circ E^{(s)} \circ \dots \circ E^{(1)}$,
where the $E^{(s)}$ are feature extractor blocks and $Q$ is a scalar differentiable quantization function (see Sec.~\ref{sec:quantization}). The $D^{(s)}$ in Fig.~\ref{fig:arch} are predictor blocks, and we parametrize $E^{(s)}$ and $D^{(s)}$ as convolutional neural networks. 

Letting $z^{(0)} = x$,
we parametrize the conditional distributions  
for all $s \in \{0, \dots, S \}$ as
\begin{equation*}
    p(z^{(s)}|z^{(s+1)}, \ldots, z^{(S)}) = p(z^{(s)}|f^{(s+1)}) \label{eq:p_cond_param},
\end{equation*}
using the predictor features
$    f^{(s)}=D^{(s)}(\fs{s+1}, z^{(s)})$.\footnote{
            The final predictor only sees $z^{(S)}$, i.e., we let $\fs{S+1} = 0$.}
Note that $\fs{s+1}$ summarizes the information of $z^{(S)}, \dots, z^{(s+1)}$. 

The predictor is based on the super-resolution architecture from EDSR~\cite{Lim_2017_CVPR_Workshops}, motivated by the fact that our prediction task is somewhat related to super-resolution in that both are dense prediction tasks involving spatial upsampling. We mirror the predictor to obtain the feature extractor, and follow~\cite{Lim_2017_CVPR_Workshops} in not using BatchNorm~\cite{ioffe2017batch}.
Inspired by the ``atrous spatial pyramid pooling'' from~\cite{chen2017rethinking}, we insert a similar layer at the end of $D^{(s)}$: In $A*$, we use three atrous convolutions in parallel, with rates 1, 2, and 4, then concatenate the resulting feature maps to a $3C_f$-dimensional feature map.

\subsection{Quantization}\label{sec:quantization}

We use the scalar quantization approach proposed in~\cite{mentzer2018cvpr} to quantize the output of $E^{(s)}$: Given levels $\levels = \{\level_1, \dots, \level_L\} \subset \R$, we use nearest neighbor assignments to quantize each entry $z' \in z^{(s)}$ as
\begin{equation} \label{eq:hardquant}
z = Q(\zunquantized) := \text{arg min}_{j} \|\zunquantized-\level_j\|,
\end{equation}
but use differentiable ``soft quantization''
\begin{equation} \label{eq:softquant}
\tilde Q(\zunquantized) = \sum_{j=1}^L  \frac{\exp(-\sigmaq\|\zunquantized-\level_j\|)}{\sum_{l=1}^L \exp(-\sigmaq\|\zunquantized-\level_l\|)} \, \level_j
\end{equation}
to compute gradients for the backward pass, where $\sigmaq$ is a hyperparameter relating to the ``softness'' of the quantization. 
For simplicity, we fix $\levels$ to be $L=25$ evenly spaced values in $[-1, 1]$.

\subsection{Mixture Model} \label{sec:mixture_model}

For ease of notation, let $z^{(0)}=x$ again. We model the conditional distributions
$p(z^{(s)}|z^{(s+1)}, \ldots, z^{(S)})$
using a generalization of the discretized logistic mixture model with $K$ components proposed in~\cite{Salimans2017pcnnpp}, as it allows for efficient training: 
The alternative of predicting logits per \mbox{(sub-)pixel} has the downsides of requiring more memory, causing sparse gradients (we only get gradients for the logit corresponding to the ground-truth value), and does not model that neighbouring values in the domain of $p$ should have similar probability. 

Let $c$ denote the channel and $u,v$ the spatial location. For all scales, we assume the entries of $z^{(s)}_{cuv}$ to be independent across $u,v$, given $f^{(s+1)}$. 
For RGB ($s=0$), we define
\begin{equation}
    p(x|f^{(1)}) = \prod_{u,v} p(x_{1uv},x_{2uv},x_{3uv}|f^{(1)}), \label{p_x_factorization}
\end{equation}
    where we use a weak autoregression over RGB channels to define the joint probability distribution via a mixture $\pmix$ (dropping the indices $uv$ for shorter notation):
\begin{align}
    p(x_1, x_2, x_3 | f^{(1)}) =\;&\pmix(x_1 | f^{(1)}) \cdot \pmix(x_2 | f^{(1)}, x_1) \; \cdot \nonumber \\
                                 &\pmix(x_3 | f^{(1)}, x_2, x_1) .
    \label{eq:p_joint}
\end{align}
    We define $\pmix$ as a mixture of logistic distributions $\plogistic$ (defined in Eq.~\eqref{eq:plog} below). To this end, we obtain mixture weights\footnote{%
    Note that in contrast to~\cite{Salimans2017pcnnpp} we do not share mixture weights $\pi^k$ across channels. This allows for easier marginalization of Eq.~\eqref{eq:p_joint}.}
    $\pi^k_{cuv}$, means $\mu^k_{cuv}$, variances $\sigma^k_{cuv}$, as well as coefficients $\lambda^k_{cuv}$ from $f^{(1)}$ (see further below), and get
\begin{align}
    \pmix(x_{1uv} | f^{(1)}) &= \sum_k \pi^k_{1uv} \; \plogistic(x_{1uv} | \tilde \mu^k_{1uv}, \sigma^k_{1uv}) \nonumber \\
    \pmix(x_{2uv} | f^{(1)}, x_{1uv}) &= \sum_k \pi^k_{2uv} \; \plogistic(x_{2uv} | \tilde \mu^k_{2uv}, \sigma^k_{2uv}) \nonumber \\
    \pmix(x_{3uv} | f^{(1)}, x_{1uv}, x_{2uv}) &= \sum_k \pi^k_{3uv} \; \plogistic(x_{3uv} | \tilde \mu^k_{3uv}, \sigma^k_{3uv}), \label{eq:p_mix}
\end{align}
    where we use the conditional dependency on previous $x_{cuv}$ to obtain the updated means $\tilde \mu$, as in~\cite[Sec. 2.2]{Salimans2017pcnnpp},
\begin{align}
    \tilde \mu^k_{1uv} = \mu^k_{1uv} \nonumber \hspace{5em}
    &\tilde \mu^k_{2uv} = \mu^k_{2uv} + \lambda^k_{\alpha uv} \; x_{1uv} \nonumber  \\
    \tilde \mu^k_{3uv} = \mu^k_{3uv} + \lambda^k_{\beta uv} &\; x_{1uv} + \lambda^k_{\gamma uv} \; x_{2uv}.
    \label{eq:updating_mus}
\end{align}
Note that the autoregression over channels in Eq.~\eqref{eq:p_joint} is only used to update the means $\mu$ to $\tilde \mu$.

For the other scales ($s>0$), the formulation only changes in that we use no autoregression at all, i.e., $\tilde \mu_{cuv} = \mu_{cuv}$ for all $c, u, v$. 
No conditioning on previous channels is needed, and Eqs.~\eqref{p_x_factorization}-\eqref{eq:p_mix} simplify to
\begin{align}
    p(z^{(s)} | f^{(s+1)}) &= \prod_{c,u,v} \pmix(z^{(s)}_{cuv} | f^{(s+1)}) \\
    \pmix(z^{(s)}_{cuv} | f^{(s+1)}) &= \sum_k \pi^k_{cuv} \; \plogistic(x_{cuv} | \mu^k_{cuv}, \sigma^k_{cuv}). \label{eq:p_mix_others}
\end{align}

For all scales, the individual logistics $\plogistic$ are given as 
\begin{align}
    \plogistic(z | \mu, \sigma) = \bigl(%
        &\sigmoid((z + b/2 - \mu) / \sigma) - \nonumber \\%
        &\sigmoid((z - b/2 - \mu) / \sigma) \bigr). \label{eq:plog}
\end{align}
Here, $b$ is the bin width of the quantization grid ($b=1$ for $s=0$ and $b=1/12$ otherwise). The edge-cases $z=0$ and $z=255$ occurring for $s=0$ are handled as described in \cite[Sec. 2.1]{Salimans2017pcnnpp}.

For all scales, we obtain the parameters of $p(z^{(s-1)}|f^{(s)})$ from $f^{(s)}$ with a $1{\times}1$ convolution that has $C_p^{(s-1)}$ output channels (see Fig.~\ref{fig:arch_detail}).
For RGB, this final feature map must contain the three parameters $\pi, \mu, \sigma$ for each of the 3 RGB channels and $K$ mixtures, as well as $\lambda_\alpha, \lambda_\beta, \lambda_\gamma$ for every mixture, thus requiring $C_p^{(0)} = 3 \cdot 3 \cdot K + 3 \cdot K$  channels. For $s>0$, $C_p^{(s)} = 3 \cdot C \cdot K$, since no $\lambda$ are needed.
With the parameters, we can obtain $p(z^{(s)}|f^{(s+1)})$, which has dimensions $3{\times}H{\times}W{\times}256$ for RGB and $C{\times}H'{\times}W'{\times}L$ otherwise (visualized with cubes in Fig.~\ref{fig:arch}).

We emphasize that in contrast to~\cite{Salimans2017pcnnpp}, our model is \emph{not} autoregressive over pixels, i.e., $z_{cuv}^{(s)}$ are modelled as independent across $u, v$ given $f^{(s+1)}$ (also for $z^{(0)}=x$).

\begin{table*}[ht!]
\centering
    \begin{tabular}{ll@{\hskip 6ex}r@{\hskip 1mm}l@{\hskip 8ex}r@{\hskip 1mm}l@{\hskip 8ex}r@{\hskip 1mm}l}
\toprule
        \footnotesize{[bpsp]} & Method & \multicolumn{2}{@{}l}{Open Images} & \multicolumn{2}{@{}l}{DIV2K}  & \multicolumn{2}{@{}l}{\raisek}  \\
\midrule
Ours & L3C & 2.991 & & 3.094 & & 2.387 &\\ \midrule
\multirow{2}{*}{\shortstack[l]{Learned\\Baselines}} & RGB Shared & 4.314 & \textcolor{tablegreen}{\footnotesize $+44\%$} & 4.429 & \textcolor{tablegreen}{\footnotesize $+43\%$} & 3.779 & \textcolor{tablegreen}{\footnotesize $+58\%$}\\
 & RGB & 3.298 & \textcolor{tablegreen}{\footnotesize $+10\%$} & 3.418 & \textcolor{tablegreen}{\footnotesize $+10\%$} & 2.572 & \textcolor{tablegreen}{\footnotesize $+7.8\%$}\\ \midrule
\multirow{4}{*}{\shortstack[l]{Non-Learned\\Approaches}} & PNG & 4.005 & \textcolor{tablegreen}{\footnotesize $+34\%$} & 4.235 & \textcolor{tablegreen}{\footnotesize $+37\%$} & 3.556 & \textcolor{tablegreen}{\footnotesize $+49\%$}\\
 & JPEG2000  & 3.055 & \textcolor{tablegreen}{\footnotesize $+2.1\%$} & 3.127 & \textcolor{tablegreen}{\footnotesize $+1.1\%$} & 2.465 & \textcolor{tablegreen}{\footnotesize $+3.3\%$}\\
 & WebP & 3.047 & \textcolor{tablegreen}{\footnotesize $+1.9\%$} & 3.176 & \textcolor{tablegreen}{\footnotesize $+2.7\%$} & 2.461 & \textcolor{tablegreen}{\footnotesize $+3.1\%$}\\
 & FLIF & 2.867 & \textcolor{tablered}{\footnotesize $-4.1\%$} & 2.911 & \textcolor{tablered}{\footnotesize $-5.9\%$} & 2.084 & \textcolor{tablered}{\footnotesize $-13\%$}\\ \bottomrule
\end{tabular}
\caption{\label{table:results_bpsp}Compression performance of our method (\name) and learned baselines (RGB Shared and RGB) to previous (non-learned) approaches, in bits per sub-pixel (bpsp). We emphasize the difference in percentage to our method for each other method in \emph{\textcolor{tablegreen}{green}} if \name outperforms the other method and in \emph{\textcolor{tablered}{red}} otherwise. \emph{Numbers are updated, see suppl.~\ref{sec:version3} for details.}}
\vspace{-1.5ex}
\end{table*}

\subsection{Loss}

We are now ready to define the loss, which is a generalization of the discrete logistic mixture loss introduced in~\cite{Salimans2017pcnnpp}. Recall from Sec.~\ref{sec:losslesscomp} that our goal is to model the true joint distribution of $x$ and the representations $z^{(s)}$, i.e., $\tilde p(x, z^{(1)},\ldots,z^{(s)})$ as accurately as possible using our model $p(x, z^{(1)},\ldots,z^{(s)})$. Thereby, the $z^{(s)}=F^{(s)}(x)$ are defined using the learned feature extractor blocks $E^{(s)}$, and $p(x, z^{(1)},\ldots,z^{(s)})$ is a product of discretized (conditional) logistic mixture models with parameters defined through the $f^{(s)}$, which are in turn computed using the learned predictor blocks $D^{(s)}$. As discussed in Sec.~\ref{sec:losslesscomp}, the expected coding cost incurred by coding $x, z^{(1)},\ldots,z^{(s)}$ w.r.t.\ our model $p(x, z^{(1)},\ldots,z^{(s)})$ is the cross entropy $H(\tilde p,p)$.

We therefore directly minimize $H(\tilde p,p)$ w.r.t.\ the parameters of the feature extractor blocks $E^{(s)}$ and predictor blocks $D^{(s)}$ over samples. Specifically, given $N$ training samples $x_1, \dots, x_N$, let $\lossparts{s} = F^{(s)}(x_\lossidx)$ be the feature representation of the $i$-th sample.
We minimize
\begin{align}
    &\mathcal L(E^{(1)}, \ldots, E^{(S)}, D^{(1)}, \ldots, D^{(S)}) \nonumber\\
    &=-\sum_{\lossidx=1}^N \log 
        \Big(
            p\big(x_\lossidx, \lossparts{1}, \ldots, \lossparts{S}\big)
        \Big) \nonumber\\
    &=-\sum_{\lossidx=1}^N \log 
        \Big(
            p\big(x_\lossidx|\lossparts{1}, \ldots, \lossparts{S}\big) \nonumber\\[-2ex]
    &\hspace{5em} 
                  \boldsymbol{\cdot} \prod_{s=1}^{S} p
                              \big(\lossparts{s}|\lossparts{s+1}, \ldots, \lossparts{S}\big) 
        \Big) \nonumber\\[-1ex]
     &=-\sum_{\lossidx=1}^N \Big(\log p(x_\lossidx|\lossparts{1}, \ldots, \lossparts{S}) \nonumber\\[-2ex]
     & \hspace{3.5em}+ \sum_{s=1}^{S}\log p(\lossparts{s}|\lossparts{s+1}, \ldots, \lossparts{S})\Big)\label{eq:coding_cost}.
\end{align}

Note that the loss decomposes into the sum of the cross-entropies of the different representations. Also note that this loss corresponds to the negative log-likelihood of the data w.r.t.\ our model which is typically the perspective taken in the generative modeling literature (see, e.g., \cite{van2016pixel}).

\paragraph{Propagating Gradients through Targets}
We emphasize that in contrast to the generative model literature, we learn the representations, propagating gradients to both $E^{(s)}$ and $D^{(s)}$, since each component of our loss depends on $D^{(s+1)}, \dots, D^{(S)}$ via the parametrization of the logistic distribution and on $E^{(s)}, \dots, E^{(1)}$ because of the differentiable $Q$. 
Thereby, our network can autonomously learn to navigate the trade-off between a) making the output $z^{(s)}$ of feature extractor $E^{(s)}$ more easily estimable for the predictor $D^{(s+1)}$ and b) putting enough information into $z^{(s)}$ for the predictor $D^{(s)}$ to predict $z^{(s-1)}$.

\subsection{Relationship to MS-PixelCNN} \label{sec:mspcnn}
When the auxiliary features $z^{(s)}$ in our approach are restricted to a non-learned RGB pyramid (see baselines in Sec.~\ref{sec:experiments}), this is somewhat similar to MS-PixelCNN~\cite{reed2017parallel}.
In particular, \cite{reed2017parallel} combines such a pyramid with upscaling networks which play the same role as the predictors in our architecture. Crucially however, they rely on combining such predictors with a shallow PixelCNN and upscaling one dimension at a time ($W{\times}H{\rightarrow}2W{\times}H{\rightarrow}2W{\times} 2H$). While their complexity is reduced from $O(WH)$ forward passes needed for PixelCNN~\cite{van2016pixel} to $O(\log W H)$, their approach is in practice still two orders of magnitude slower than ours (see Sec.~\ref{sec:results_pixelcnn}).
Further, we stress that these similarities only apply for our RGB baseline model, whereas our best models are obtained using learned feature extractors trained jointly with the predictors.

\section{Experiments} \label{sec:experiments}

\paragraph{Models} We compare our main model (\name) to two learned baselines:
For the \emph{RGB Shared} baseline (see Fig.~\ref{fig:arch_rgb_shared}) we use bicubic subsampling as feature extractors, i.e., $z^{(s)}=\Bicubic_{2^s}(x)$, and only train one predictor $D^{(1)}$. During testing, we obtain multiple $z^{(s)}$ using $\Bicubic$ and apply the single predictor $D^{(1)}$ to each.
The \emph{RGB} baseline (see Fig.~\ref{fig:arch_rgb}) also uses bicubic subsampling, however, we train $S=3$ predictors $D^{(s)}$, one for each scale, to capture the different distributions of different RGB scales.
For our main model, \emph{\name}, we additionally learn $S=3$ feature extractors $E^{(s)}$.\footnote{We chose $S=3$ because increasing $S$ comes at the cost of slower training, while yielding negligible improvements in bitrate. For an image of size $H{\times}W$, the last bottleneck has $5{\times}H/8{\times}W/8$ dimensions, each quantized to $L=25$ values. Encoding this with a uniform prior amounts to ${\approx}4\%$ of the total bitrate. For the RGB Shared baseline, we apply $D^{(1)}$ 4 times, as only one encoder is trained.}
Note that the only difference to the RGB baseline is that the representations $z^{(s)}$ are learned. We train all these models until they converge at 700k iterations. 

\paragraph{Datasets}
We train our models on 362\,551 images randomly selected from the \emph{Open Images} training dataset~\cite{OpenImages}.
We randomly downscale the images to at least 512 pixels on the longer side to remove potential artifacts from previous compression, discarding images where rescaling does not result in at least $1.25\times$ downscaling.
Further, following~\cite{balle2018variational} we discard high saturation/ non-photographic images, i.e., images with mean $S{>}0.9$ or $V{>}0.8$ in the HSV color space.
We evaluate on 500 images randomly selected from the Open Images validation set, 
preprocessed like the training set (available on our github\textsuperscript{\ref{fn:github}}), the 100 images from the commonly used super-resolution dataset \emph{DIV2K}~\cite{agustssondiv2k},
as well as on \emph{\raisek}~\cite{dang2015raise}, a ``real-world image dataset'' with 1000 images. We automatically split images into equally-sized crops if they do not fit into GPU memory,
and process crops sequentially. Note that this is a bias \emph{against} our method.

In order to compare to the PixelCNN literature, we additionally train \name on the ImageNet32 and ImageNet64 datasets~\cite{chrabaszcz2017downsampled},
each containing  1\,281\,151 training images and 50\,000 validation images, of $32 \times 32$ resp. $64 \times 64$ pixels.

\paragraph{Training} We use the RMSProp optimizer~\cite{hinton2012neural}, with a batch size of 30, minimizing Eq.~\eqref{eq:coding_cost} directly (no regularization). 
We train on $128 \times 128$ random crops, and apply random horizontal flips.
We start with a learning rate $\lambda=1\cdot10^{-4}$ and decay it by a factor of 0.75 every 5 epochs. 
On ImageNet32/64, we decay $\lambda$ every epoch, due to the smaller images.

\paragraph{Architecture Ablations} We find that adding BatchNorm~\cite{ioffe2015batch} slightly degrades performance. Furthermore, replacing the stacked atrous convolutions $A*$ with a single convolution, slightly degrades performance as well. By stopping gradients from propagating through the targets of our loss, we get significantly worse performance -- in fact, the optimizer does not manage to pull down the cross-entropy of any of the learned representations $z^{(s)}$ significantly.

We find the choice of $\sigmaq$ for $Q$ has impacts on training:~\cite{mentzer2018cvpr} suggests setting it s.t.\ $\tilde Q$ resembles identity, which we found to be good starting point, but found it beneficial to let $\sigmaq$ be slightly smoother (this yields better gradients for the encoder). We use $\sigmaq=2$.

Additionally, we explored the impact of varying $C$ (number of channels of $z^{(s)}$) and the number of levels $L$ and found it more beneficial to increase $L$ instead of increasing $C$, i.e., it is beneficial for training to have a finer quantization grid.

\paragraph{Other Codecs} We compare to FLIF and the lossless mode of WebP using the respective official implementations~\cite{flif2016, webpurl}, for PNG we use the implementation of Pillow~\cite{pillowurl},
and for the lossless mode of \jpegk we use the Kakadu implementation~\cite{kakaduurl}.
See Sec.~\ref{sec:related_work} for a description of these codecs.

\section{Results}

\begin{table}
\centering
\begin{tabular}{p{1ex}lll} \toprule
    & Method & $32\times32$px & $320 \times 320$px \\
 \midrule
    \multirow{2}{*}[0.25ex]{%
    \rotatebox[origin=c]{90}{%
        \small BS=1}} & \name (Ours)    
            & 0.0168 s &    0.0291 s          \\
                    & PixelCNN++~\cite{Salimans2017pcnnpp}
            & 47.4 s\tablenote{*} &     $\approx 80$ min\tablenote{\ddagger} \\
\midrule
    \multirow{4}{*}[-0.5ex]{%
    \rotatebox[origin=c]{90}{%
            \centering\small  BS=30}} & \name (Ours)    
            & 0.000624 s &    0.0213 s           \\
& PixelCNN++
            & 11.3 s\tablenote{*} &       $\approx18$ min\tablenote{\ddagger}  \\ 
\cmidrule(l){2-4}
&PixelCNN~\cite{van2016pixel}
            & 120 s\tablenote{\dagger}  &      $\approx 8$ hours\tablenote{\ddagger} \\
&MS-PixelCNN~\cite{reed2017parallel}
            & 1.17 s\tablenote{\dagger} &      $\approx 2$ min\tablenote{\ddagger}   \\
\bottomrule 
\end{tabular}
\caption{
    \label{table:times}Sampling times for our method (\name), compared to the PixelCNN literature. The results in the first two rows were obtained with batch size (BS) 1, the other times with BS=30, since this is what is reported in~\cite{reed2017parallel}. 
        \deftablenote{*}: Times obtained by us with code released of PixelCNN++~\cite{Salimans2017pcnnpp}, on the same GPU we used to evaluate \name (Titan X Pascal). 
        \deftablenote{\dagger}: times reported in~\cite{reed2017parallel}, obtained on a Nvidia Quadro M4000 GPU (no code available).
        \;\deftablenote{\ddagger}: To put the numbers into perspective, we compare our runtime with \emph{linearly extrapolated} runtimes for for the other approaches  on $320 \times 320$ crops.
    }
\end{table}

\subsection{Compression} \label{sec:results_compression}

Table~\ref{table:results_bpsp} shows a comparison of our approach (\name) and the learned baselines to the other codecs, on our testsets, in terms of bits per sub-pixel (bpsp)\footnote{We follow the likelihood-based generative modelling literature in measuring bpsp; $X$ bits per pixel (bpp) $=X/3$ bpsp, see also footnote~\ref{fn:subpixel}.}
All of our methods outperform the widely-used PNG, which is at least $43\%$ larger on all datasets. We also outperform WebP and \jpegk everywhere by a smaller margin of up to $3.3\%$.
We note that FLIF still marginally outperforms our model but remind the reader of the many hand-engineered highly specialized techniques involved in FLIF (see Section~\ref{sec:related_work}). In contrast, we use a simple convolutional feed-forward neural network architecture. 
The RGB baseline with $S=3$ learned predictors outperforms the RGB Shared baseline on all datasets, showing the importance of learning a predictor for each scale.
Using our main model (\name), where we 
additionally learn the feature extractors, we outperform both baselines: The outputs are at least $7.8\%$ larger everywhere, showing the benefits of learning the representation.

\begin{table}[t]
\centering
    \begin{tabular}{lcc} 
        \toprule
        \footnotesize{[bpsp]} & ImageNet32 &   Learned \\
        \midrule
        \name (ours) & 4.76 &  \checkmark \\
        PixelCNN~\cite{van2016pixel} & 3.83 &  \checkmark    \\
        MS-PixelCNN~\cite{reed2017parallel} & 3.95 &  \checkmark  \\
        \midrule
        PNG & 6.42 \\
        \jpegk& 6.35 \\
        WebP & 5.28 \\
        FLIF & 5.08 \\
        \bottomrule
\end{tabular}
    \caption{\label{table:results_imgnet32}Comparing bits per sub-pixel (bpsp) on the $32 \times 32$ images from ImageNet32 of our method (\name) vs.\ PixelCNN-based approaches and classical approaches.}
\end{table}

\subsection{Comparison with PixelCNN} \label{sec:results_pixelcnn}
While PixelCNN-based approaches are not designed for lossless image compression, they learn a probability distribution over pixels and optimize for the same log-likelihood objective. Since they thus can in principle be used inside a compression algorithm, we show a comparison here.

\paragraph{Sampling Runtimes} Table~\ref{table:times} shows a speed comparison to three PixelCNN-based approaches (see Sec.~\ref{sec:related_work} for details on these approaches).
We compare time spent when \emph{sampling} from the model, to be able to compare to the PixelCNN literature. Actual decoding times for \name are given in Sec.~\ref{sec:decoding_time}.

While the runtime for PixelCNN~\cite{van2016pixel} and MS-PixelCNN~\cite{reed2017parallel} is taken from the table in~\cite{reed2017parallel}, we can compare with \name by assuming that PixelCNN++ is not slower than PixelCNN to get a conservative estimate\footnote{PixelCNN++ is in fact around $3\times$ faster than PixelCNN due to modelling the joint directly, see Sec.~\ref{sec:related_work}.},
and by considering that MS-PixelCNN reports a $105{\times}$ speedup over PixelCNN.
When comparing on $320 \times 320$ crops, we thus observe massive speedups compared to the original PixelCNN: ${>}1.63\cdot10^5\times$ for batch size (BS) 1 and ${>}5.31\cdot10^4\times$ for BS 30. 
We see that on $320 \times 320$ crops, \name is \emph{at least} $5.06\cdot10^2\times$ faster than MS-PixelCNN, the fastest PixelCNN-type approach.
Furthermore, Table~\ref{table:times} makes it obvious that the PixelCNN based approaches are not practical for lossless compression of high-resolution images. 

We emphasize that it is impossible to do a completely fair comparison with PixelCNN and MS-PixelCNN due to the unavailability of their code and the different hardware. Even if the same hardware was available to us, differences in frameworks/framework versions (PyTorch vs.\ Tensorflow) can not be accounted for. See also Sec.~\ref{sec:note_pcnn_cmp} for notes on the influence of the batch size.

\paragraph{Bitcost} To put the runtimes reported in Table~\ref{table:times} into perspective, we also evaluate the bitcost on ImageNet32, for which PixelCNN and MS-PixelCNN were trained, in Table~\ref{table:results_imgnet32}.
We observe our outputs to be $20.6\%$ larger than MS-PixelCNN and $24.4\%$ larger than the original PixelCNN, but smaller than all classical approaches. However, as shown above, this increase in bitcost is traded against orders of magnitude in speed. We obtain similar results for ImageNet64, see Sec.~\ref{sec:imgnet64_cmp}.

\subsection{Encoding / Decoding Time} \label{sec:decoding_time}

\addtolength{\tabcolsep}{-0.8ex}    
\begin{table}[b]
    \begin{tabular}{lllcc@{\hskip 1mm}c}
        \toprule
        Codec & Encoding [s] & Decoding [s] & [bpsp] & GPU & CPU \\
        \midrule
        \name (Ours) & 0.242 & 0.374 & $3.386$ & \checkmark & \checkmark \\
        \midrule
        PNG & $0.213$ & $6.09\cdot10^{-5}$ & $4.733$ &&\checkmark \\
        \jpegk & $1.48\cdot10^{-2}$ & $2.26\cdot10^{-4}$ & $3.471$  &&\checkmark\\
        WebP & $0.157$ & $7.12\cdot10^{-2}$ & $3.447$  &&\checkmark\\
        FLIF & $1.72$ & $0.133$ & $3.291$   &&\checkmark\\
        \bottomrule
    \end{tabular}
    \caption{\label{table:enc_dec_vs_classical}Encoding and Decoding times compared to classical approaches, on $512 \times 512$ crops from DIV2K, as well as bpsp and required devices.}
\end{table}
\addtolength{\tabcolsep}{0.9ex}    

To encode/decode images with \name (and other methods outputting a probability distribution), a pass with an entropy coder is needed. We implemented a relatively simple pipeline to encode and decode images with \name, which we describe in the supplementary material, in Section~\ref{sec:suppl_decoding_details}. The results are shown in Tables~\ref{table:enc_dec_vs_classical} and \ref{table:enc_dec_details}. As noted in Section~\ref{sec:suppl_decoding_details}, we did not optimize our code for speed, yet still obtain practical runtimes.
We also note that to use other likelihood-based methods for lossless compression, similar steps are required.
While our encoding time is in the same order as for classical approaches, our decoder is slower than that of the other approaches. This can be attributed to more optimized code and offloading complexity to the encoder -- while in our approach, decoding essentially mirrors encoding. However, combining encoding and decoding time we are either faster (FLIF) or have better bitrate (PNG, WebP, \jpegk).

\begin{figure}[t]
\centering
    \footnotesize
    \setlength{\tabcolsep}{0.5mm}
    \begin{tabular}{lrlr}
    \multicolumn{2}{l}{\includegraphics[width=0.49\linewidth]{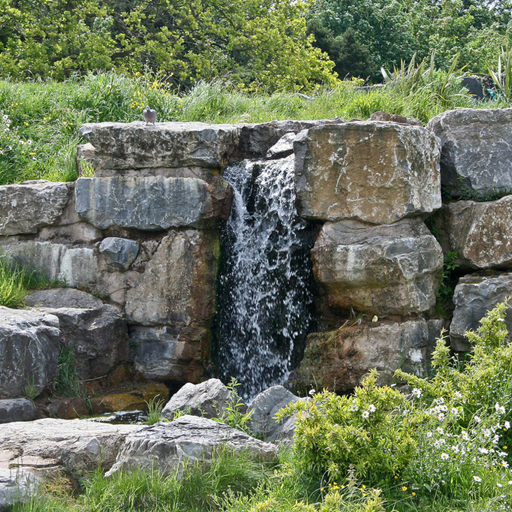}} &
    \multicolumn{2}{r}{\includegraphics[width=0.49\linewidth]{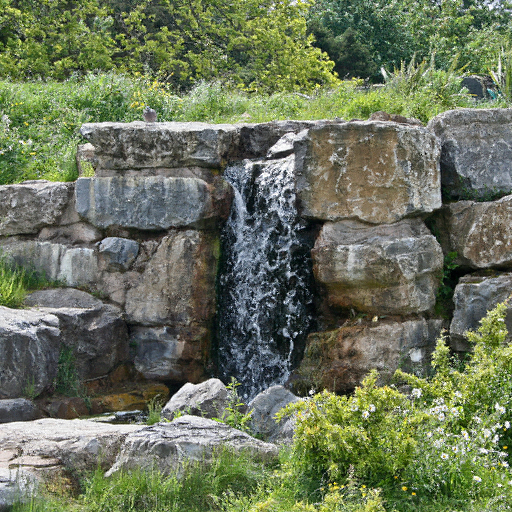}} \\[-0.8ex]
    5.207 bpsp &  \emph{stored: 0,1,2,3} &
    1.209 bpsp &    \emph{stored: 1,2,3} \\
    \multicolumn{2}{l}{\includegraphics[width=0.49\linewidth]{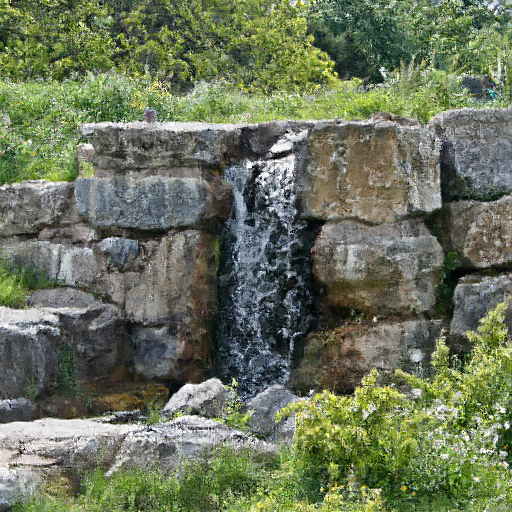}} &
    \multicolumn{2}{r}{\includegraphics[width=0.49\linewidth]{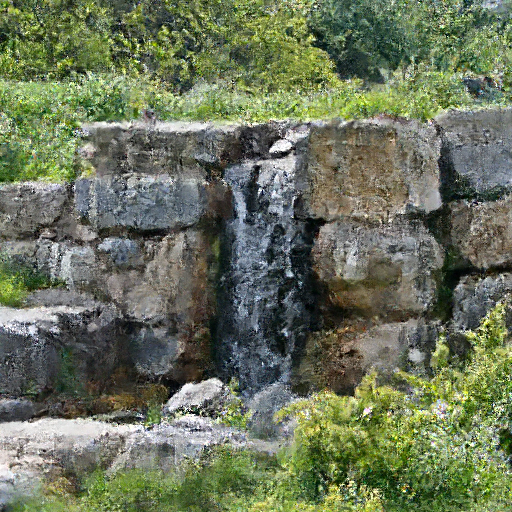}} \\[-0.8ex]
    0.355 bpsp &     \emph{stored: 2,3} &
    0.121 bpsp &       \emph{stored: 3} \\
\end{tabular}
    \caption{\label{fig:generation}Effect of generating representations instead of storing them, given different $z^{(s)}$ of a $512 \times 512$ image. Below each generated image, we show the required bitcost and which scales are stored.}
\end{figure}

\subsection{Sampling Representations}
We stress that we study image compression and not image generation. Nevertheless, our method produces models from which $x$ and $z^{(s)}$ can be sampled. Therefore, we visualize the output when sampling part of the representations from our model in
Fig.~\ref{fig:generation}: the top left shows an image from Open Images, when we store all scales (losslessly). When we store  $z^{(1)}, z^{(2)}, z^{(3)}$ but not $x$ and instead sample from $p(x|f^{(1)})$, we only need $27.8\%$ of the total bits without noticeably degrading visual quality. Sampling $z^{(1)}$ and $x$ leads to some blur while reducing the number of stored bits to $10.2\%$ of the full bitcost. Finally, only storing $z^{(3)}$ (containing $64 \times 64 \times 5$ values from $\levels$ and $3.00\%$ of the full bitcost) and sampling $z^{(2)}$, $z^{(1)}$, and $x$ produces significant artifacts. However, the original image is still recognizable, showing the ability of our networks to learn a hierarchical representation capturing global image structure.

\vspace{-0.3ex}\section{Conclusion}\vspace{-0.2ex}

We proposed and evaluated a fully parallel hierarchical probabilistic model with auxiliary feature representations. 
Our \name model outperforms PNG, \jpegk and WebP on all datasets. Furthermore, it significantly outperforms the RGB Shared and RGB baselines which rely on predefined heuristic feature representations, showing that learning the representations is crucial. Additionally, we observed that using PixelCNN-based methods for losslessly compressing full resolution images takes two to five orders of magnitude longer than \name.

To further improve \name, future work could investigate weak forms of autoregression across pixels and/or dynamic adaptation of the model network to the current image. Moreover, it would be interesting to explore domain-specific applications, e.g., for medical image data.

\begin{spacing}{0.8}
\vfill
{\setlength{\parindent}{0cm}\footnotesize\selectfont
\textbf{Acknowledgments} The authors would like to thank Sergi Caelles for the insightful discussions and feedback.
This work was partly supported by ETH General Fund (OK) and Nvidia through the hardware grant.}
\end{spacing}

{\small
\bibliographystyle{ieee}
\bibliography{egbib}
}

\clearpage

\appendix

\setcounter{table}{0}
\renewcommand{\thetable}{A\arabic{table}}
\setcounter{figure}{0}
\renewcommand{\thefigure}{A\arabic{figure}}

\twocolumn[\section{Practical Full Resolution Learned Lossless Image Compression -- Supplementary Material}\vspace{2em}]

\subsection{Changes in Version 3} \label{sec:version3}

Previously, our preprocessing script saved all training images and validation images as JPGs with a high quality factor of $Q=95$, downscaled by a factor 0.75. It turns out that the resulting images have a specific enough distribution that the neural network picks up on it, and the images are also easier to compress for the non-learned codecs. 

For correctness, we have thus re-created the training and validation sets. The new preprocessing script and more details is available on github\textsuperscript{\ref{fn:github}}. The important differences are:
\begin{itemize}[leftmargin=*,topsep=0pt,itemsep=-1ex,partopsep=1ex,parsep=1ex]
\item All images are saved as PNGs.
\item We do not rescale validation sets in any way, and instead divide the images into crops such that everything fits into memory.
\item For the training set, we use a random downscaling factor, instead of fixed 0.75x: this provides a wider variety of downscaling artefacts.
\end{itemize}

\subsection{Encoding and Decoding Details} \label{sec:suppl_decoding_details}

\begin{table}[b]
    \centering
    \begin{tabular}{lll}
        \toprule
         Decoding Time & Obtaining CDF      & Arithmetic \\
                       & for Decoder [s]     & Decoding [s] \\
         \midrule
         $s=3, 64\times64$      & - & 0.00179 \\
         $s=2, 128\times128$    & 0.00737 & 0.00759 \\
         $s=1, 256\times256$    & 0.0219 & 0.0234 \\
         $s=0, 512\times512$    & 0.143 & 0.169 \\
         \midrule
         Total & 0.172 & 0.202 \\
         \bottomrule
    \end{tabular}
    \caption{\label{table:enc_dec_details}We show the time to obtain CDF, including all forward passes through the different stages, as well as the time required by the arithmetic decoder. We measured on a Titan X (Pascal), and took the average over 500 crops of $512 \times 512$ pixels. For $s=3$, we assume a uniform prior, and thus do not need to calculate a CDF.}
\end{table}

Table~\ref{table:enc_dec_details} shows the time required to decode each scale $s$. We first obtain the CDF as a matrix on the CPU to be able to use the arithmetic decoder (see below), and then do a pass with the arithmetic decoder. We did not optimize either part for speed, as noted in Sec.~\ref{sec:notes_optim}. 

The following shows detailed steps, using again $z^{(0)} = x$. The steps are also visualized in Fig.~\ref{fig:encdecdetails}.

\subsubsection*{Encoding}
\begin{enumerate}[leftmargin=*]
    \item Forward pass through network to obtain $\forall s: \; z^{(s)}$, $f^{(s)}$.
    \item Encode $z^{(S)}$ assuming a uniform prior, i.e., assuming each of the $L$ symbols is equally likely. This requires $\log_2(L)$ bits per symbol.
    \item Update the means $\mu$ predicted for the RGB scale ($s=0$) to $\tilde \mu$, given the input $x$ (see Eq.~\eqref{eq:updating_mus}).
    \item In practice, the division into intervals $[a,b)$ required for arithmetic coding described in Sec.~\ref{sec:losslesscomp} is most efficiently done by having access to the cumulative distribution function (CDF) of the symbol to encode. 
        Thus, for the RGB scale ($s=0$), we obtain the CDF analogously to Eq.~\eqref{eq:p_mix}:
        \begin{align}
            \hspace{-2ex}C(x_{1uv} | f^{(1)}) &= \sum_k \pi^k_{1uv} \; C_l(x_{1uv} | \tilde \mu^k_{1uv}, \sigma^k_{1uv}) \nonumber \\
            \hspace{-2ex}C(x_{2uv} | f^{(1)}, x_{1uv}) &= \sum_k \pi^k_{2uv} \; C_l(x_{2uv} | \tilde \mu^k_{2uv}, \sigma^k_{2uv}) \nonumber \\
            \hspace{-2ex}C(x_{3uv} | f^{(1)}, x_{1uv}, x_{2uv}) &= \sum_k \pi^k_{3uv} \; C_l(x_{3uv} | \tilde \mu^k_{3uv}, \sigma^k_{3uv}). \label{eq:C_mix}
\end{align}
        And, analogously to Eq.~\eqref{eq:p_mix_others}, the CDF for $s>0$ for each channel $c$ is
        \begin{equation}
            C(z^{(s)}_{cuv} | f^{(s+1)}) = \sum_k \pi^k_{cuv} C_l(z^{(s)}_c | \mu^k_{cuv}, \sigma^k_{cuv}).
            \label{eq:cdf_z_s_c}
        \end{equation}
        $C_l$ in Eqs.~\eqref{eq:C_mix},~\eqref{eq:cdf_z_s_c} is the CDF of the logistic distribution, \[ C_l(z | \mu, \sigma) = \sigmoid((z-\mu)/\sigma). \]
        For each $s, c$ the CDF $C(z^{(s)}_{c} | f^{(s+1)})$ is a $H' \times W' \times L$-dimensional matrix, where $L=257$ for RGB and $L=26$ otherwise, and $H' = H/2^s, W' = W/2^s$. 
    \item For each $s \in \{S+1, \dots, 0\}$, encode each channel $c$ of $z^{(s)}$ with the predicted $C(z^{(s)}_{c} | f^{(s+1)})$, using adaptive arithmetic coding (see Sec.~\ref{sec:losslesscomp}). To be able to uniquely decode, the sub-bitstream for $z^{(s)}$ always starts with a triplet encoding its dimensions $C, H', W'$ as \textsc{uint16}. The final bitstream is the concatenation of all sub-bitstreams.
\end{enumerate}

\subsubsection*{Decoding}
\begin{enumerate}[leftmargin=*]
    \item Obtain the final $z^{(S)}$ from the bitstream, which was encoded with a uniform prior.
    \item Feed $z^{(S)}$ to $D^{(S)}$ to obtain $f^{(S)}$, and thereby also $C(z^{(S-1)}_{cuv} | f^{(S)})$ for all $c$. Since the decoder now has access to the same CDF as the encoder, we can decode $z^{(S-1)}$ from the bitstream with our adaptive arithmetic decoder.
    \item Analogously, we repeat the previous step to obtain $z^{(S)}, \dots, z^{(1)}$, as well as $f^{(S)}, \dots, f^{(1)}$ using the accompanying CDFs.
    \item Given $f^{(1)}$, which contains all parameters for the RGB scale (i.e., we know $\forall k, c, u, v$: $\pi^k_{cuv}, \mu^k_{cuv}, \sigma^k_{cuv}$ as well as $\lambda^k_{\alpha uv}, \lambda^k_{\alpha uv}, \lambda^k_{\alpha uv}$, see Sec.~\ref{sec:mixture_model}), we can obtain the CDF for the first channel of $x$ ($x_1$, red channel), $C(x_1|f^{(1)})$, and decode this first channel from the bitstream. Now we know $x_1$, and with $\mu^k_{2uv}, \lambda_{\alpha uv}^k$ we can obtain $\tilde \mu^k_2$ via Eq.~\eqref{eq:updating_mus}. With this, we also know the CDF of the next channel, $C(x_2|f^{(1)}, x_1)$, and can decode $x_2$ from the bitstream. In the same fashion, we can then obtain $\tilde \mu_3^k$, then $C(x_3|f^{(1)}, x_1, x_2)$, and thus $x_3$.
    \item Concatenating the channels $x_1, x_2, x_3$, we finally obtain the decoded image $x$.
\end{enumerate}

\subsubsection{Hardware Used}

Our timings were obtained on a machine with a Titan X (Pascal) GPU and Intel Xeon E5-2680 v3 CPU.

\subsubsection{Notes on Code Optimization} \label{sec:notes_optim}

The encoder can be run in parallel over all scales, as all CDFs are known after one forward pass. Further, we do not need to know the CDF for all symbols, but only for the symbols $z$ we encode and $z+1$, since this specifies the interval $[a, b)$. 
The decoder is sequential in the scales since $z^{(s)}$ is required to predict the distribution of $z^{(s-1)}$. Still, for $s>0$, the decoding of the channels of the $z^{(s)}$ could be parallelized, as the channels are modelled fully independently.
However, we did not implement either of these improvements, keeping the code simple.

For both encoder and decoder, the CDFs must be available to the CPU, as the arithmetic coder runs there. However, the CDFs are huge tensors for real-world images ($H \times W \times 257$ for RGB, which amounts to $257$MB for each channel of a $512 \times 512$ image). To save the expensive copying from GPU to CPU, we implemented our own CUDA kernel to store the claculated $C$ directly into ``managed memory'', which can be accessed from both CPU and GPU. However, we did not optimize this CUDA kernel for speed.

Finally, while state-of-the-art adaptive entropy coders typically require on the order of milliseconds per MB (see \cite{duda2015use} and in particular \cite{giesen2014interleaved} for benchmarks on adaptive entropy coding), we implemented a simple arithmetic coding module to obtain the times in our tables.
Please see the code\textsuperscript{\ref{fn:github}} for details.

\newpage

\subsection{Comparison on ImageNet64} \label{sec:imgnet64_cmp}

We show a bpsp comparison on ImageNet64 in Table~\ref{table:results_imgnet64}. Similar to what we observed on ImageNet32 (see Section~\ref{sec:results_pixelcnn}), our outputs are $23.8\%$ larger than MS-PixelCNN and $19.4\%$ larger than the original PixelCNN, but smaller than all classical approaches. We note again that increase in bitcost is traded against orders of magnitude in speed. 

We also note that the gap between classical approaches and PixelCNN becomes smaller compared to ImageNet32. 

\begin{table}[ht!]
    \centering
    \begin{tabular}{lcc} 
        \toprule
        \footnotesize{[bpsp]} & ImageNet64 &   Learned \\
        \midrule
        \name (ours) & 4.42 &  \checkmark \\
        PixelCNN~\cite{van2016pixel} & 3.57 &  \checkmark    \\
        MS-PixelCNN~\cite{reed2017parallel} & 3.70 &  \checkmark  \\
        \midrule
        PNG & 5.74 \\
        \jpegk & 5.07 \\
        WebP & 4.64 \\
        FLIF & 4.54 \\
        \bottomrule
\end{tabular}
    \captionsetup{width=.8\linewidth}
    \caption{\label{table:results_imgnet64}Comparing bits per sub-pixel (bpsp) on the $64 \times 64$ images from ImageNet64 of our method (\name) vs.\ PixelCNN-based approaches and classical approaches.}
\end{table}

\subsection{Note on Comparing Times for $32 \times 32$ Images} \label{sec:note_pcnn_cmp}
In Table~\ref{table:times}, we report run times for batch size 30 to be able to compare with the run times reported in~\cite{reed2017parallel}. However, this comparison is biased against us, as can be seen in Table~\ref{table:batchsize_32}: Since our network is fairly small, we can process up to 480 images of size $32{\times}32$ in parallel. We observe that the time to sample one image drops as the batch size increases, indicating that for BS=30, some overhead dominates.

\begin{table}[h!]
    \centering
    \begin{tabular}{p{0.39\linewidth}l}
        \toprule
        Batch Size & Time per image [s] \\
        \midrule
        30 & $6.24\cdot10^{-4}$ \\
        60 & $4.31\cdot10^{-4}$ \\
        120 & $3.16\cdot10^{-4}$ \\
        240 & $2.52\cdot10^{-4}$ \\
        480 & $2.42\cdot10^{-4}$ \\
        \bottomrule
    \end{tabular}
    \captionsetup{width=.8\linewidth}
    \caption{\label{table:batchsize_32}Effect of varying the batch size.\hfill}
\end{table}

\clearpage

\subsection{Visualizing Representations} \label{sec:visualize_repr}

\begin{figure}[ht]
\centering
    \includegraphics[width=\linewidth]{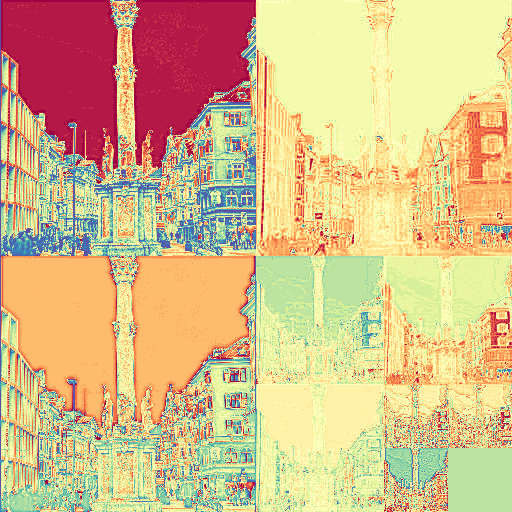}\vspace{-0.9ex}
    \includegraphics[width=\linewidth]{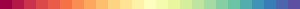}\vspace{-3.1ex}
    {\tiny \textcolor{white}{1\hfill25}}
    \caption{\label{fig:bn}Heatmap visualization of the first three channels for each of the representations $z^{(1)}, z^{(2)}, z^{(3)}$, each containing values in $\levels=\{1,\dots,25\}$, as indicated by the scale underneath.}
\end{figure}

We visualize the representations $z^{(1)}, z^{(2)}, z^{(3)}$ in Fig.~\ref{fig:bn}. It can be seen that the global image structure is preserved over scales, with representations corresponding to smaller $s$ modeling more detail. This shows potential for efficiently performing image understanding tasks on partially decoded images similarly as described in \cite{torfason2018towards} for lossy learned compression: instead of training a feature extractor for a given task on $x$, one could directly use the features $z^{(s)}$ from our network.

\newpage

\subsection{Architectures of Baselines}
\begin{figure}[ht]
\centering
\includegraphics[width=0.8\linewidth]{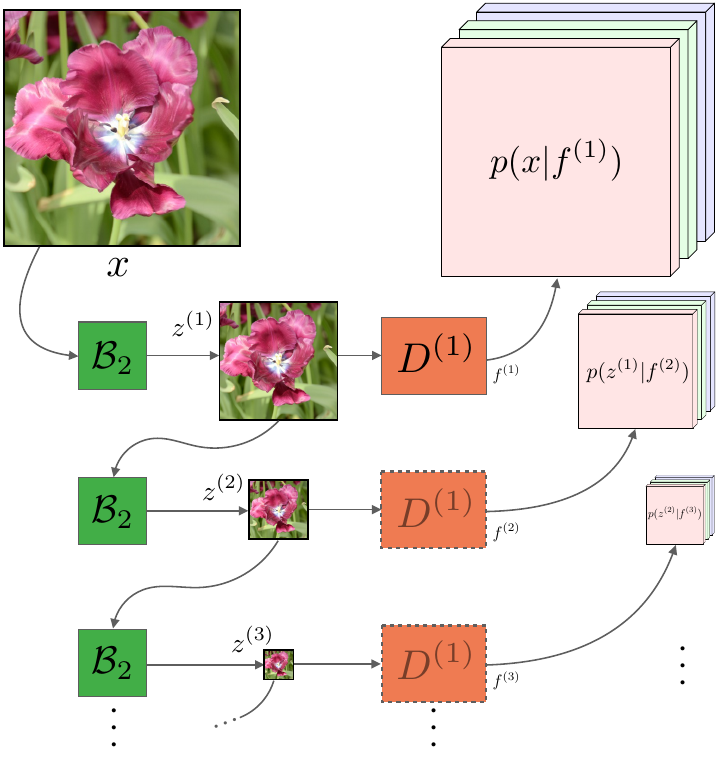}
    \caption{\label{fig:arch_rgb_shared}Architecture for the RGB Shared baseline. Note that we train only one predictor $D^{(1)}$.}
\end{figure}

\begin{figure}[ht!]
\centering
\includegraphics[width=0.8\linewidth]{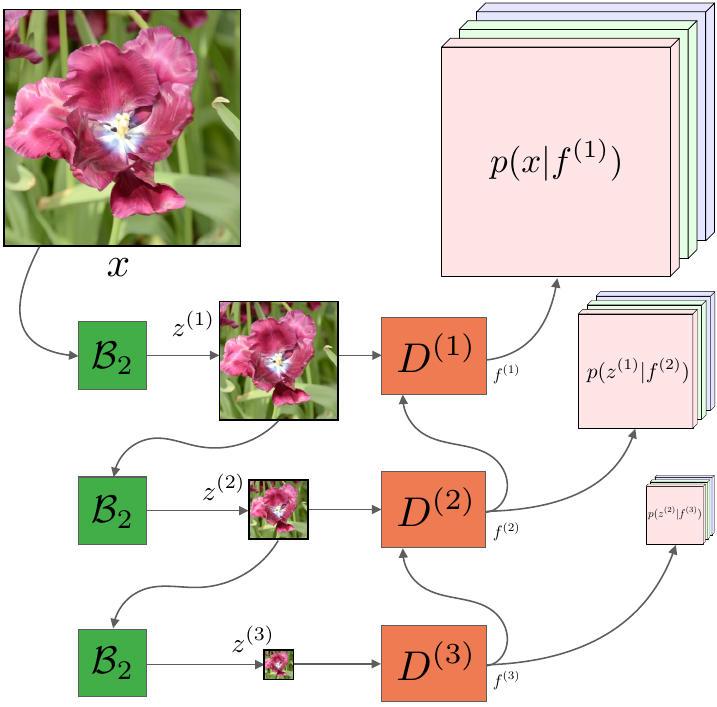}
    \caption{\label{fig:arch_rgb}Architecture for the RGB baseline. Multiple predictors are trained.}
\end{figure}

Figs.~\ref{fig:arch_rgb_shared},~\ref{fig:arch_rgb} show the architectures for the RGB Shared and RGB baselines. The dots in Fig.~\ref{fig:arch_rgb_shared} indicate that the model could in theory be applied more since $D^{(1)}$ is used for every scale.

\subsection{Encoding and Decoding Visualized}
\begin{figure*}[hb]
\centering
    \vspace{-2em}
    Encoding\hspace{2em} Decoding\\
\includegraphics[width=0.8\textwidth]{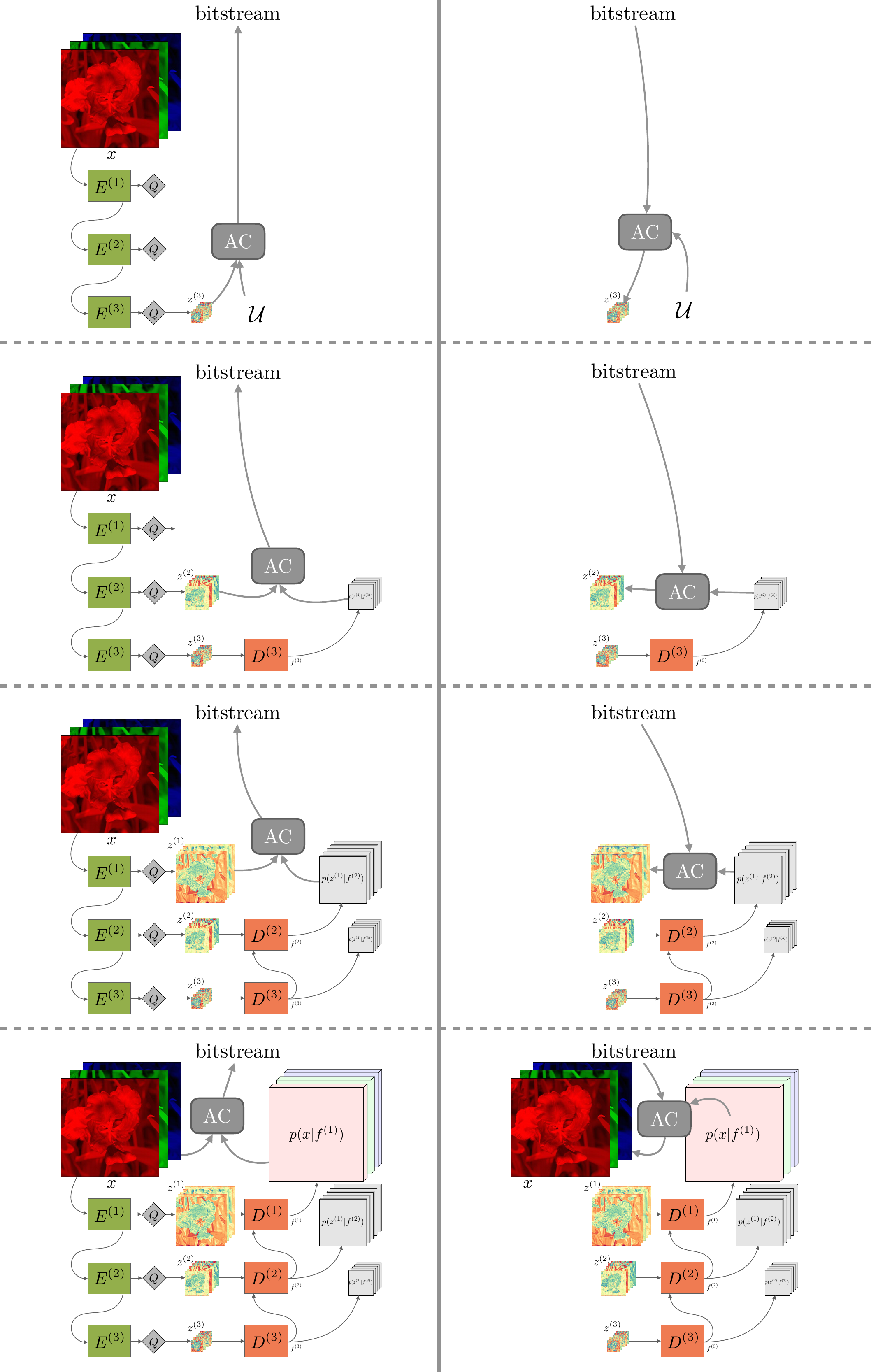}
    \caption{\label{fig:encdecdetails}Visualizing encoding and decoding: At every step, the arithmetic coder (AC) takes a probability distribution and a $z^{(s)}$.} 
\end{figure*}

We visualize the steps needed to encode the different $z^{(s)}$ in Fig.~\ref{fig:encdecdetails} on the next page.

\end{document}